\documentclass[10pt,twoside]{article}
\usepackage{amsmath}
\usepackage{amssymb}
\usepackage{appendix}
\usepackage{graphicx}
\usepackage{caption}
\usepackage{subcaption}
\usepackage{authblk}
\usepackage{geometry}
\title{The General Complex Envelope Solutions of Coupled Mode Optics with Quadratic or Cubic Nonlinearity}
\author{Graham D. Hesketh: G.Hesketh@soton.ac.uk}
\affil{Optoelectronics Research Centre University of Southampton, Highfield, Southampton, Hampshire, SO17 1BJ, United Kingdom}

\begin{document}

\maketitle
\begin{abstract}
The analytic general solutions for the complex field envelopes are derived using Weierstrass elliptic functions for two and three mode systems of differential equations coupled via quadratic $\chi_2$ type nonlinearity as well as two mode systems coupled via cubic $\chi_3$ type nonlinearity. For the first time, a compact form of the solutions is given involving simple ratios of Weierstrass sigma functions (or equivalently Jacobi theta functions). A Fourier series is also given. All possible launch states are considered. The models describe sum and difference frequency generation, polarization dynamics, parity-time dynamics and optical processing applications. 

\copyright 2015 Optical Society of America. One print or electronic copy may be made for personal use only. Systematic reproduction and distribution, duplication of any material in this paper for a fee or for commercial purposes, or modifications of the content of this paper are prohibited. Published here: https://www.osapublishing.org/josab/abstract.cfm?uri=josab-32-12-2391
\end{abstract}

\section{Introduction}\label{bintro}

Coupled mode nonlinear optical systems form a fundamental building block in optical processing functionality. Herein, we provide an extensive analytical study of both the coupled mode system for quadratic nonlinearity, found in materials such as periodically poled lithium niobate (PPLN) crystals and commonly referred to as $\chi_2$ in optics, as well as the coupled mode system for cubic nonlinearity commonly referred to as $\chi_3$ and typically found in silica waveguides such as optical fibers. The functionality enabled by the quadratic nonlinearity includes, but is not limited to, sum and difference frequency generation \cite{Armstrong1962, Goodman2015}, modulation format conversion \cite{Ros2014}, optical logic \cite{Zhang2011,Jiang2015}, and all-optical switching \cite{Ironside1993, Hutchings1993, Kobyakov1996, Asobe1997}. Respectively, the cubic nonlinear system in optics can describe nonlinear polarization mode dynamics; which include instabilities \cite{Winful1986} and has enabled such functionality as the optical Kerr shutter \cite{Duguay1969,Agrawal2007} and all-optical regeneration \cite{Parmigiani2014,Parmigiani2015}, adjacent waveguides coupled via evanescent optical fields; which are utilized in all-optical switching and optical logic \cite{Jensen1982}, and most recently, nonlocal non-Hermitian parity-time (PT) symmetric coupled mode systems; which may provide a test bed within optics for theoretical physics and possibly enable novel unique optical properties such as non-reciprocity between modes and power threshold behaviour (\textit{see}  \cite{Sarma2014} and references therein). Outside of optics, the coupled mode equations with cubic nonlinearity also describe Bose-Einstein condensates (BECs) in atomic physics \cite{Ostrovskaya2000}. 

The equations describing light interactions in a dielectric with quadratic or cubic nonlinearity were first derived by Armstrong et. al. in 1962 \cite{Armstrong1962}. Therein, the amplitudes of two or three mixing frequencies were solved for in terms of Jacobi elliptic functions. In a Hamiltonian analysis, bifurcations and instabilities were examined in the two mode quadratic nonlinear system in \cite{Trillo1993} and the phase of the fundamental mode was solved for as an elliptic integral of the third kind in \cite{Ironside1993} under the special condition that the second mode (or second harmonic) was zero at input. It was proposed therein that the intensity dependent refractive index could be used to induce phase shifts in a Mach-Zehnder waveguide to produce an all-optical switch. A similar technique was explored in a 3 mode case in \cite{Hutchings1993} where in addition to optical switching, wavelength demultiplexing applications were identified. But it was in \cite{Kobyakov1996}, that the 3 mode exact solutions for the amplitudes and phases of modes coupled under a quadratic nonlinearity were given in terms of elliptic functions and integrals, respectively, for general launch conditions. In that work, which again considered optical switching applications, the modes were considered to be different frequencies and different polarizations, although from a mathematical perspective that case is a rescaling of the 3 mode case in \cite{Armstrong1962}. 

In the cubic nonlinear case, the amplitudes of modes in two adjacent waveguides coupled via their evanescent fields were solved for in terms of elliptic functions while investigating optical logic applications in \cite{Jensen1982}, while in \cite{Peng1990}, elliptic integral expressions for the phases of the two modes were derived and analysed. The amplitudes of two polarization modes coupled via cubic nonlinearity in the circular polarization basis were solved for in terms of Jacobi elliptic functions in \cite{Winful1986} while investigating polarization instabilities. In \cite{Khare2015}, special solutions in the form of Jacobi elliptic functions have also been found in the continuous version of the PT model introduced in \cite{Sarma2014}, although solutions to the discrete coupled mode counterpart have to our knowledge not yet been provided. In addition to these relevant works, there is also a notable amount of literature which studies both the quadratic and cubic models in both Hamiltonian and geometric formalisms (\textit{see} e.g \cite{Holm2008} and references therein). 

Herein, we study the 2 and 3 mode system of equations coupled under quadratic nonlinearity and the 2 mode system coupled under cubic nonlinearity. Importantly however, our paper differs to the existing literature we have discussed in three main ways. It is known that the nonlinear differential equations describing power evolution in the optical models under consideration are typically solved by elliptic functions. However, as will be discussed, the differential equations involve cubic rather than quartic polynomials in the power evolution and this makes the Weierstrass elliptic function notation the natural choice as the power in each mode is then linear in the Weierstrass function rather than bi-linear or bi-quadratic in the Jacobi function \cite{WhitWat1915}. Some researchers also consider Weierstrass elliptic notation to be more modern and it is thus useful to have the choice. Firstly then, we present the power evolution in the Weierstrass notation unlike any of the existing literature which favours the Jacobi notation. Secondly, and more importantly, we go on to obtain our main result, the general solution to the complex representation of the electric field envelope. These expressions have never been derived in any of the previous works. Our result is expressible in a surprisingly simple way as the ratio of two Weierstrass sigma functions \cite{WhitWat1915} (or alternatively Jacobi theta functions) and this simplicity is in contrast to the somewhat more complicated separate expressions obtained in the literature for the power flow and (where it is obtained) the phase. Notably, the method used herein can incorporate arbitrary launch conditions. While for the general complex envelope this requires the inversion of the Weierstrass function, which can be done numerically, in the case of the power flow this inversion can be entirely mitigated if one implements addition identities of Weierstrass elliptic functions as is discussed. This may be preferred in physical problems if one wishes to better display parameter dependence. Thirdly and finally, we recognise that our form of the solution is closely related to a Kronecker double series \cite{Weil1976} and this enables us to give a Fourier series form of the solution. This has also never been done before. It is envisioned that this form will prove particularly useful given how important Fourier analysis is in optics and it is important that these solutions are obtained for completeness as they themselves can often yield new approximations and insights. As a test, the analytic solutions are plotted against numerical solutions obtained with a standard Runge Kutta technique and found to be in exact agreement. All elliptic function theory used herein can be found in \cite{WhitWat1915}, although some important identities are reproduced in the appendices for convenience.

The layout of the paper is as follows. Firstly, in section \ref{bquad} we tackle the quadratic case. From the literature we introduce the model of 3 modes coupled via quadratic nonlinearity. The general solution is then found in terms of sigma functions and an analytic example is presented against a numerical solution as a test. This system is then reduced to a degenerate special cases of two modes and another solved example is presented. In section \ref{bcube} we tackle the cubic case. We introduce a generalized model of four independent modes coupled via cubic nonlinearity and the general solution is then found in terms of sigma functions. This system is then reduced to special cases of two modes, plus their complex conjugates, and models from the literature are considered as examples, including the Hermitian example of two polarization modes \cite{Agrawal2007} and the non-Hermitian case of a parity-time symmetric system \cite{Sarma2014}. In section \ref{wideapps} we discuss possible wider applications of the solution method.

\section{Quadratic Case}\label{bquad}

In this section we tackle the quadratic case. Let us consider the following 3 mode equations which are, up to a phase rotation and rescaling of the fields and length variable $z$, equivalent to those appearing in \cite{Armstrong1962,Hutchings1993,Kobyakov1996,Asobe1997,Smith1995} (\textit{see} \ref{AppTrans} for how to transform from a conventional set of equations with material parameters in physical units to the normalised form in \eqref{EQ31}):
\begin{equation}\label{EQ31}
\begin{aligned}
A_{1}'&=i A_{1}+iA_2^*A_{3}\\
A_{2}'&=i A_{2}+iA_1^*A_{3}\\
A_{3}'&=i A_{3}+i2A_{1}A_{2}\\
\end{aligned}
\end{equation}
where $A_j$ is a complex representation of the electric field envelope, the prime denotes differentiation with respect to the length variable $z$ and the $*$ denotes complex conjugation. This system conserves the following quantities:
\begin{equation}\label{EQ32}
\begin{aligned}
P=&\left|A_{1}\right|^2+\left|A_{2}\right|^2+\left|A_{3}\right|^2,\quad Q=\left|A_{1}\right|^2-\left|A_{2}\right|^2\\
K=&\left|A_{1}\right|^2+\left|A_{2}\right|^2+\frac{1}{2}\left|A_{3}\right|^2+A_{1}A_{2}A_3^*+A_1^*A_2^*A_{3}
\end{aligned}
\end{equation}
To solve the system in \eqref{EQ31}, we first construct the differential equation for $|A_1|^2=A_1A_1^*$ which is given by:
\begin{equation}\label{EQp1}
\begin{aligned}
\left(|A_1|^2\right)'=-i\left(A_1A_2A_3^*-A_1^*A_2^*A_3\right)
\end{aligned}
\end{equation}
Then, by squaring both sides of \eqref{EQp1} and substituting in \eqref{EQ32} we can eliminate the other modes and make the right hand side a polynomial in $|A_1|^2$ as follows:
\begin{equation}\label{EQp2}
\begin{aligned}
\left[\left(|A_1|^2\right)'\right]^2=&-\left(A_1A_2A_3^*-A_1^*A_2^*A_3\right)^2\\
=&-\left(A_1A_2A_3^*+A_1^*A_2^*A_3\right)^2+4|A_1|^2|A_2|^2|A_3|^2\\
=&-8\,\left(|A_1|^2\right)^{3}- \left( 1-12Q-4P \right) \left(|A_1|^2\right)^{2}\\
&+\left(2K-P+Q-4QP-4Q^2 \right) \left(|A_1|^2\right)\\
&- \left( K-\frac{1}{2}P+\frac{1}{2}Q\right) ^{2}
\end{aligned}
\end{equation}
Differential equations like \eqref{EQp2} are solved by elliptic functions. However, because the polynomial is cubic, in this instance the Weierstrass notation is the natural choice as the defining differential equation of the Weierstrass elliptic function also involves a cubic polynomial, while that of the Jacobi function is quartic. To make \eqref{EQp2} resemble the Weierstrass elliptic differential equation more closely, let us define the following function and constants:
\begin{equation}\label{EQ33}
\begin{aligned}
\psi(z)&=p_1-2\left|A_{1}(z)\right|^2\\
p_1&=\frac{P}{3}-\frac{1}{12}+Q\\
p_2&=\frac{P}{3}-\frac{1}{12}-Q\\
p_3&=2K-P+Q\\
\end{aligned}
\end{equation}
then \eqref{EQp2} becomes:
\begin{equation}\label{EQ34}
\begin{aligned}
\left(\psi'\right)^2&=4\psi^3-g_2\psi-g_3\\
g_{{2}}&=4\left(\,{p_{{1}}}^{2}+p_{{1}}p_{{2}}+{p_{{2}}}^{2}\right)
+2p_{{3}}-p_{{1}}+p_{{2}}\\
g_3&=\left(p_{{1}}-p_{{3}}\right)^2-4\left(p_{{1}}{p_{{2}}}^{2}+{p_{{1}}}^{2}p_{{2}}\right)-p_{{1}}p_{{2}}
\end{aligned}
\end{equation}
and hence that $\psi$ is a Weierstrass elliptic function $\wp$ formed with elliptic invariants $g_2$ and $g_3$ \cite{WhitWat1915} such that:
\begin{equation}\label{EQ35}
\begin{aligned}
&\psi(z)=\wp(z-z_0)\\
&\left\{z_0\in\mathbb{C}:\wp(z_0)=\psi(0),\,\wp'(z_0)=-\psi'(0)\right\}
\end{aligned}
\end{equation}
where $\wp'$ is the derivative of the Weierstrass $\wp$ function. Let us also define the points $\xi_{j}\in\mathbb{C}$ such that:
\begin{equation}\label{EQ36}
\begin{aligned}
&\left\{\xi_1:\wp(\xi_1)=p_1,\,\wp'(\xi_1)=ip_3\right\} \\
&\left\{\xi_2:\wp(\xi_2)=p_2,\,\wp'(\xi_2)=i\left(p_3-p_1+p_2\right)\right\} \\
&\left\{
       \begin{array}{ll}
                  \xi_3:&\wp(\xi_3)=-p_1-p_2-1/4\\
                  &\wp'(\xi_3)=i\left(2p_1+p_2-p_3+1/4\right)\\

        \end{array}
\right\}
\end{aligned}
\end{equation}
and it can further be shown through addition identities that $\xi_3=\xi_1+\xi_2$. The values of $z_0$ and $\xi_{j}$ can be determined numerically by inverting $\wp$ in a mathematical software package such as Maple; $\wp'$ determines the sign, and these points are uniquely defined modulo lattice periods of the doubly periodic $\wp$ function (\textit{see} \eqref{A1}). It follows from \eqref{EQ32}-\eqref{EQ36} that the power evolution is given by:
\begin{equation}\label{EQ37}
\begin{aligned}
|A_j|^2&=r_j\left[\wp(\xi_j)-\wp(z-z_0)\right]\\
\end{aligned}
\end{equation}
where $r_1=r_2=1/2,\,r_3=-1$ and $j=1-3$. While \eqref{EQ37} is more compact, it is possible to avoid having to invert $\wp$ to find $z_0$. To do this, one should expand the $\wp(z-z_0)$ term using addition identities (\textit{see} \eqref{B1}) and it is then possible to express $\wp(\xi_j),\wp(z_0)$ and $\wp(z_0)'$ in terms of the initial conditions using \eqref{EQ32}-\eqref{EQ36} to obtain a more explicit way of seeing parameter dependence. But notably, having obtained \eqref{EQ37} one can then construct the logarithmic derivatives of $A_{j}$ in terms of $\wp,\wp'$. To do this it is convenient to initially express the logarithmic derivatives of $A_{j}$ in terms of $|A_j|^2$ and its derivative $\left(|A_j|^2\right)'$ using \eqref{EQ31}, then one can use \eqref{EQ36} and \eqref{EQ37} to obtain:
\begin{equation}\label{EQ38}
\begin{aligned}
\frac{A_j'}{A_j}&=\frac{i}{2}b_j+\frac{1}{2}\frac{r_j\wp'(\xi_{j})+\left(|A_j|^2\right)'}{|A_j|^2}\\
&=\frac{i}{2}b_j+\frac{1}{2}\frac{\wp'(\xi_{j})-\wp'(z-z_0)}{\wp(\xi_{j})-\wp(z-z_0)}\\
&=\frac{i}{2}b_j+\zeta(z-z_0+\xi_j)-\zeta(z-z_0)-\zeta(\xi_j)
\end{aligned}
\end{equation}
where $b_1=b_2=1$ and $b_3=3$, $\zeta$ is the Weierstrass zeta function \cite{WhitWat1915} and the $\zeta$ form follows from known function identities (\textit{see} \eqref{B2}). \eqref{EQ38} is then easily integrated and expressed in terms of Weierstrass $\sigma$ functions using \cite{WhitWat1915}:
\begin{equation}\label{EQ9}
\begin{aligned}
\int\frac{A_j'}{A_j}dz=\log{A_j},\,\int \zeta(z)dz=\log{\sigma(z)}
\end{aligned}
\end{equation}
to yield the first main result of this paper; the general complex envelope solutions to quadratic nonlinear systems in the form:
\begin{equation}\label{EQ39}
\begin{aligned}
A_{j}(z)&=c_j\frac{\sigma(z-z_0+\xi_j)}{\sigma(z-z_0)}e^{\rho_j z}\\
\rho_j&=\frac{i}{2}b_j-\zeta(\xi_j)
\end{aligned}
\end{equation}
where the $c_j$ are the integration constants and can be fixed by the initial conditions. The $\sigma$ functions are entire functions of the argument and have the properties $\sigma(0)=0$ and $\sigma(-z)=-\sigma(z)$. The $\sigma$ functions are also expressible in terms of the Jacobi $\theta_1$ functions (\textit{see} \eqref{B4}). It is also possible to obtain a Kronecker double series \cite{Weil1976,WhitWat1915} for \eqref{EQ39} which we present in \ref{AppK}. If \eqref{EQ37} is expressed in terms of sigma functions using known function identities (\textit{see} \eqref{B3}) then one can easily define the conjugate field as follows:
\begin{equation}\label{EQ40}
\begin{aligned}
A_j^*&=\frac{|A_{j}|^2}{A_{j}}=\frac{r_j}{c_j\sigma(z_0)^2}\frac{\sigma(z-z_0-\xi_j)}{\sigma(z-z_0)}e^{-\rho_j z}\\
\end{aligned}
\end{equation}
and subsequently the phases $\phi_j$ can be found by taking logarithms of the following:
\begin{equation}\label{EQ41}
\begin{aligned}
\mathrm{exp}\left[i2\phi_j(z)\right]&=\frac{A_{j}}{A_j^*}=\frac{c_j^2\sigma(z_0)^2}{r_j}\frac{\sigma(z-z_0+\xi_j)}{\sigma(z-z_0-\xi_j)}e^{2\rho_j z}\\
\end{aligned}
\end{equation}

\begin{figure}[ht]
\centering
  \begin{subfigure}{0.49\columnwidth}
  \includegraphics[width=1\columnwidth]{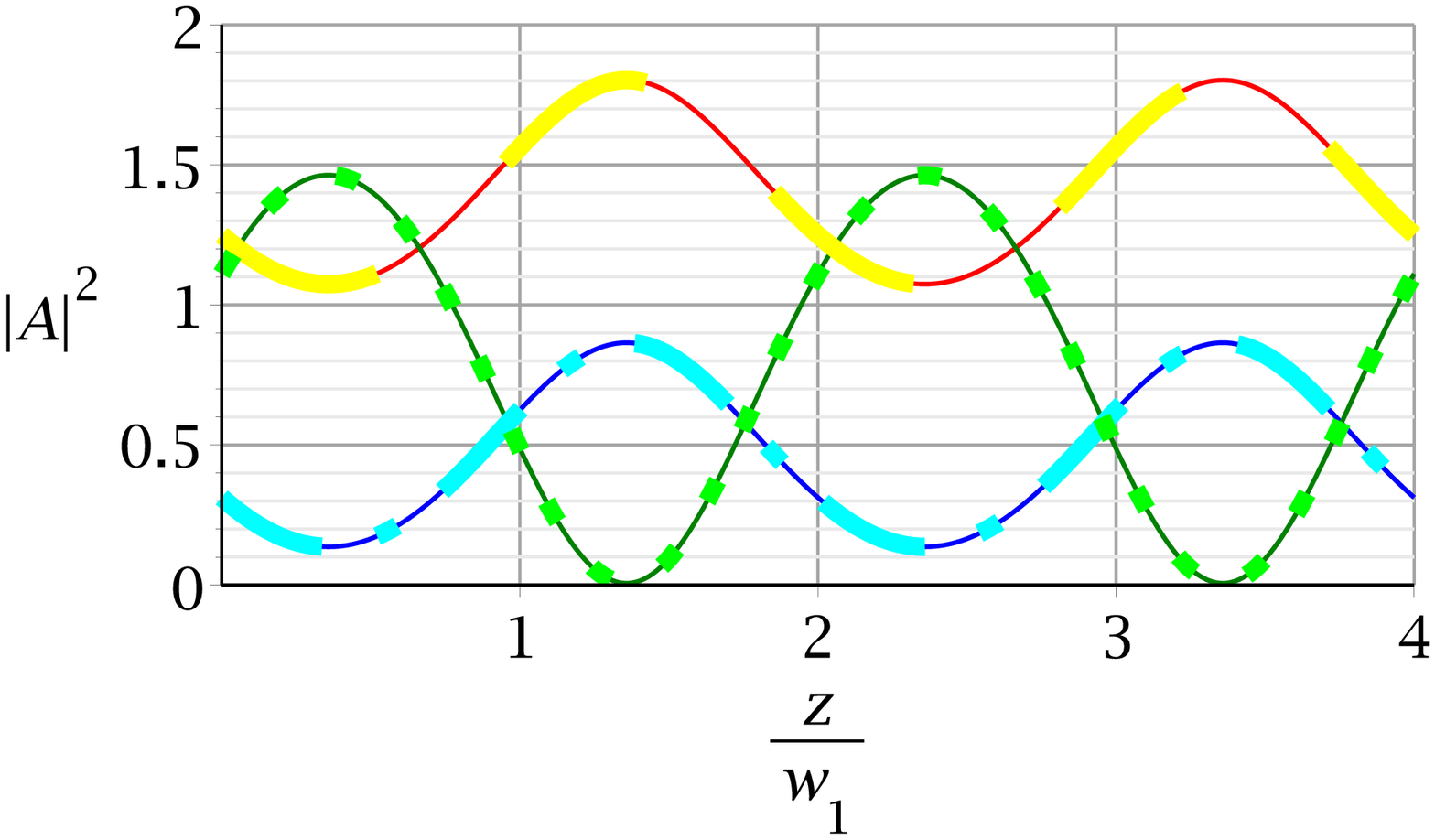}
	\caption{}
	\label{fig1a}%
	\end{subfigure}
  \begin{subfigure}{0.49\columnwidth}
  \includegraphics[width=1\columnwidth]{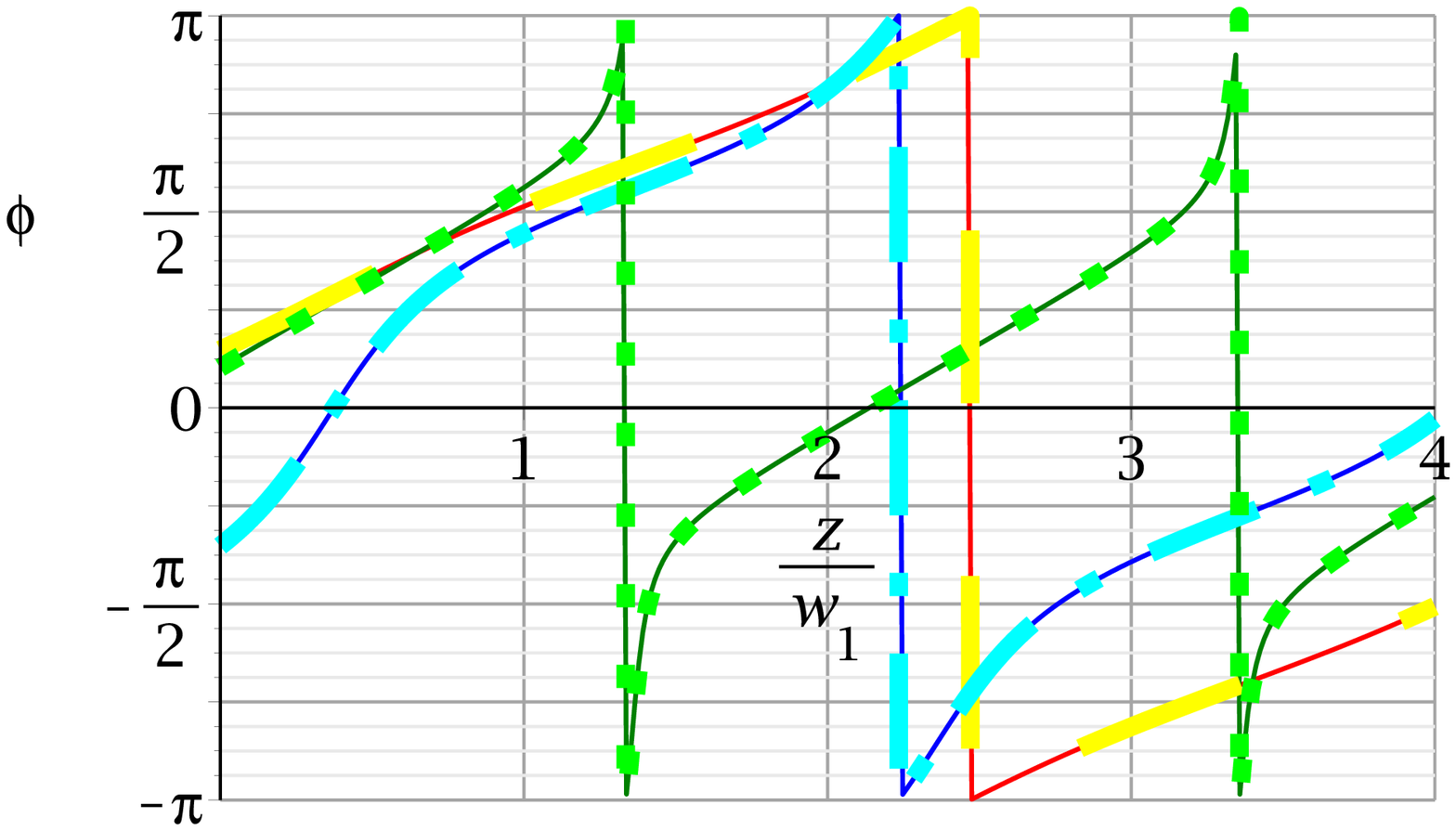}
	\caption{}
	\label{fig1b}%
	\end{subfigure}
  \caption{(a) Power $|A|^2$ and (b) phase $\phi=\text{arg}(A)$ of $A_1$ (\eqref{EQ39}: dash-yellow, Numeric: solid-red), $A_2$ (\eqref{EQ39}: dot-dash-cyan, Numeric: solid-dark blue) and $A_3$ (\eqref{EQ39}: dot-light green, Numeric: solid-dark green).}\label{fig1}
\end{figure}

Although the solutions to the phase evolution have now been solved for in \eqref{EQ41}, it is worth noting that the phases themselves evolve according to the following differential equation:
\begin{equation}\label{EQ42}
\begin{aligned}
\phi_j'(z)=-\frac{i\,r_j\,\wp'(\xi_j)}{2|A_j(z)|^2}+\frac{b_j}{2}
\end{aligned}
\end{equation}
which can be derived by writing $A_j=|A_j|e^{i\phi_j}$ in \eqref{EQ31} and using \eqref{EQ32}, \eqref{EQ33} and \eqref{EQ36}. One remarkable property of \eqref{EQ42} is that if $\wp'(\xi_j)=0$; which holds if e.g $|A_j(0)|^2=0$, then phase change is independent of intensity change with length, rather phase change comes only from the constant $b_j$ term, which in a physically scaled system could be a wavelength dispersive effect. This observation was noted and discussed in \cite{Smith1995}. As a test, Fig. \ref{fig1a} plots the magnitude and Fig. \ref{fig1b} the phase of the analytic solution given in \eqref{EQ39} for arbitrary launch conditions $A_1(0)=1+0.5\,i,A_2(0)=0.25-0.5\,i$ and $A_3(0)=1+0.33\,i$ (dash yellow, dot-dash cyan, dot light green lines, respectively) together with the equivalent numeric solutions of \eqref{EQ31} obtained using a numerical Runge Kutta method (red, dark blue and dark green solid lines, respectively). The Runge Kutta method itself was tested by confirming that the numerical solutions preserve the constants in \eqref{EQ32}. The x-axis has been expressed in units of the half-period of oscillation $w_1$ which can be obtained via a standard elliptic integral involving the invariants in \eqref{EQ34} (\textit{see} \eqref{A1}). The analytic and numeric results are indistinguishable and together with more plots not presented, this supports the validity of the solutions. Physically, the fields $A_1$ and $A_2$ might for example be considered as different frequencies of light, $f_1$ and $f_2$, propagating in a quadratic nonlinear medium and $A_3$ as their sum frequency $f_3=f_1+f_2$. In such a scenario, and in accordance with the laws of quantum mechanics, a photon is annihilated from each of frequencies $f_1$ and $f_2$ whilst two are simultaneously created at $f_3$ (or vice versa) and hence the power flows of modes $A_1$ and $A_2$ are seen to run in parallel in Fig. \ref{fig1a}, but the power in $A_3$ increases as that in $A_1$ and $A_2$ decreases (or vice versa). 

\begin{figure}[ht]
\centering
  \begin{subfigure}{0.49\columnwidth}
  \includegraphics[width=1\columnwidth]{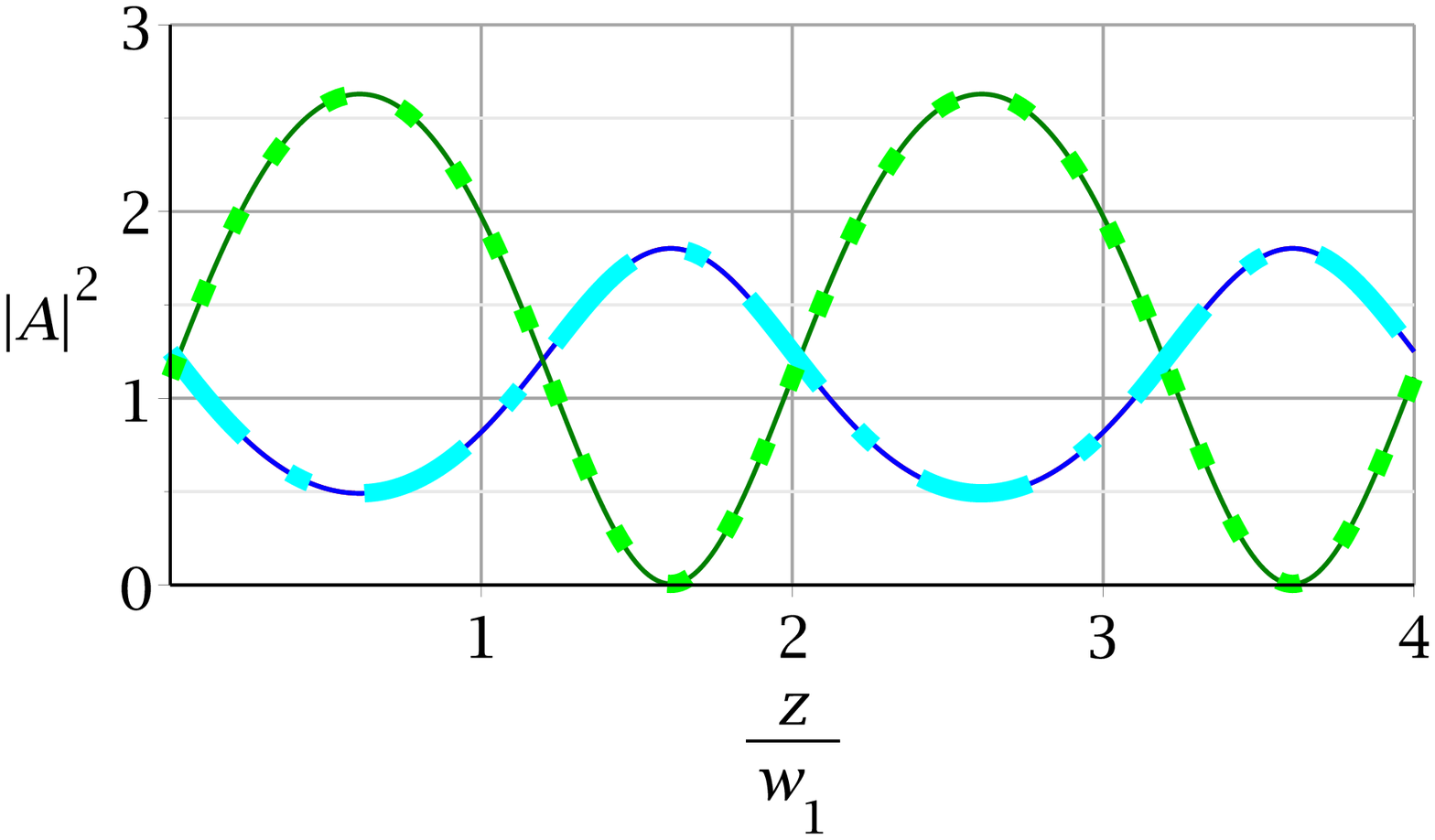}
	\caption{}
	\label{fig2a}%
	\end{subfigure}
  \begin{subfigure}{0.49\columnwidth}
  \includegraphics[width=1\columnwidth]{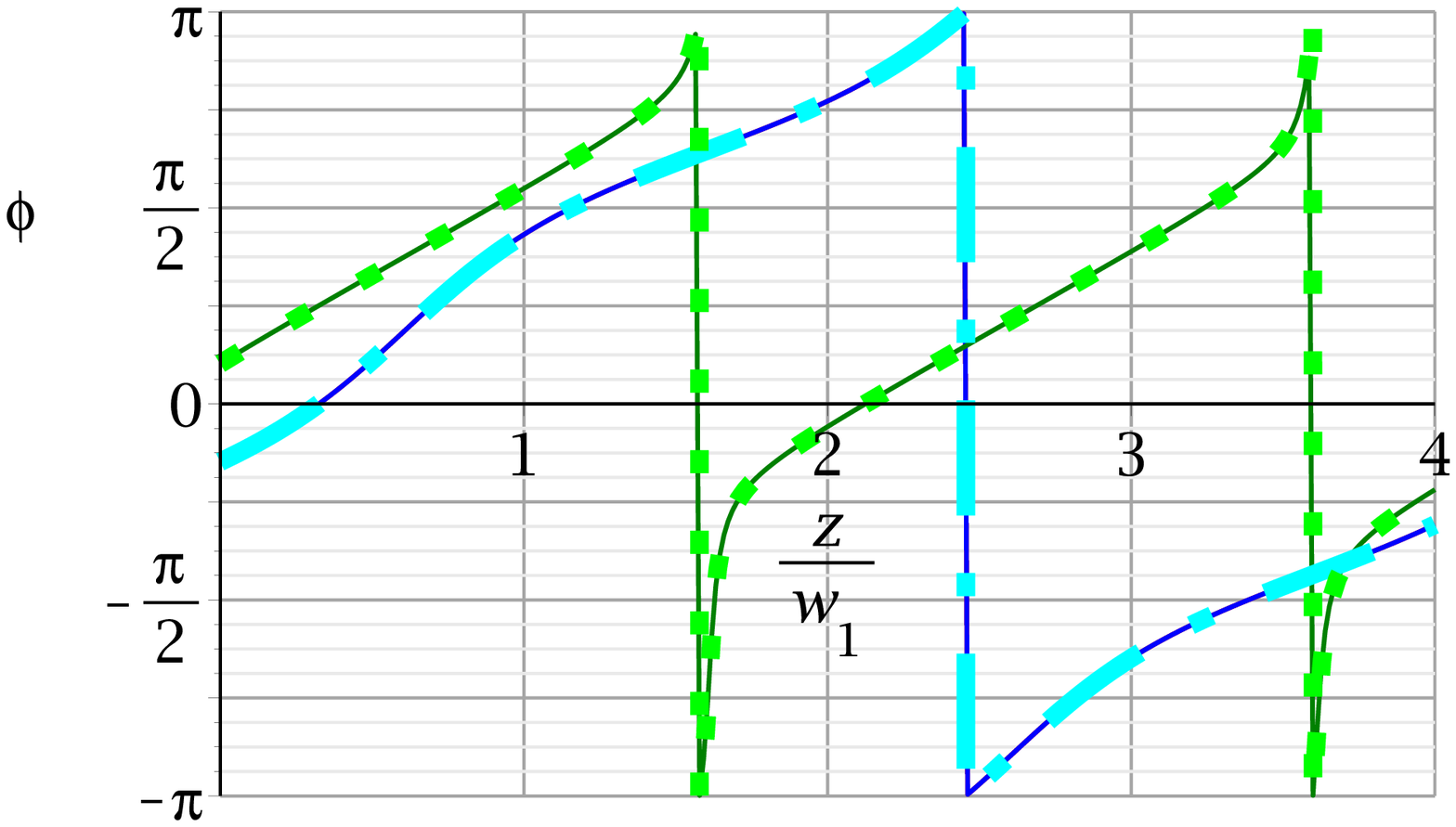}
	\caption{}
	\label{fig2b}%
	\end{subfigure}
  \caption{(a) Power $|A|^2$ and (b) phase $\phi=\text{arg}(A)$ of $A_1$ (\eqref{EQ39}: dash-dot-cyan, Numeric: solid-dark blue) and $A_3$ (\eqref{EQ39}: dot-light green, Numeric: solid-dark green).}\label{fig2}
\end{figure}

Let us now consider the important two mode case which can be viewed as a degeneracy of the three mode case if we set $A_2=A_1$ such that \eqref{EQ31} reduces to what are, up to a phase rotation and rescaling of the fields and length variable $z$, equivalent to those appearing in \cite{Armstrong1962,Goodman2015,Zhang2011,Ironside1993}:

\begin{equation}\label{EQ1}
\begin{aligned}
A_{1}'&=iA_{1}-iA_1^*A_{3}\\
A_{3}'&=iA_{3}-i2A_{1}^{2}\\
\end{aligned}
\end{equation}
where we have left the second mode to be indexed by 3 to be consistent with the notation of \eqref{EQ31}-\eqref{EQ42}. It follows that the conserved quantities in \eqref{EQ32} reduce to:
\begin{equation}\label{EQ2}
\begin{aligned}
P=&2\left|A_{1}\right|^2+\left|A_{3}\right|^2,\quad Q=0\\
K=&2\left|A_{1}\right|^2+\frac{1}{2}\left|A_{3}\right|^2+A_{1}^2A_3^*+\left(A_1^*\right)^2A_{3}\\
\end{aligned}
\end{equation}
and thus that $p_2=p_1$ in \eqref{EQ33}-\eqref{EQ34} and hence that $\xi_3=2\xi_1=2\xi_2$. Together with these reductions, \eqref{EQ39}-\eqref{EQ41} once again give the general solutions only with $j=1$ or $j=3$. Of course it would be possible to rescale such that $A_1\rightarrow A_1/\sqrt{2}$ in order to have $P=|A_1|^2+|A_3|^2$ representing total power conservation but this was avoided here in the interest of reusing notation from \eqref{EQ31}-\eqref{EQ42}. As a test, Fig. \ref{fig2a} plots the magnitude and Fig. \ref{fig2b} the phase of the analytic solution given in \eqref{EQ39} (dot-dash cyan and dot light green lines, respectively) for two modes with arbitrary launch conditions $A_1(0)=1-0.5\,i$ and $A_3(0)=1+0.33\,i$ together with the equivalent numeric solutions of \eqref{EQ31} obtained using a numerical Runge Kutta method (dark green and dark blue solid lines, respectively). The results are indistinguishable and together with more plots not presented, this supports the validity of the solutions.  

\section{Cubic Case}\label{bcube}

In this section we tackle the cubic case. Let us consider the following 4 mode equations coupled under a cubic nonlinearity:
\begin{equation}\label{bEQ31}
\begin{aligned}
A_{1}'&=i A_{1}-\frac{i}{2}\left(A_{1}^2+A_{2}^2\right)A_{3}-iA_{1}A_{2}A_{4}\\
A_{2}'&=-i A_{2}-\frac{i}{2}\left(A_{1}^2+A_{2}^2\right)A_{4}-iA_{1}A_{2}A_{3}\\
A_{3}'&=-i A_{3}+\frac{i}{2}\left(A_{3}^2+A_{4}^2\right)A_{1}+iA_{2}A_{3}A_{4}\\
A_{4}'&=i A_{4}+\frac{i}{2}\left(A_{3}^2+A_{4}^2\right)A_{2}-iA_{1}A_{3}A_{4}\\
\end{aligned}
\end{equation}
where $A_j$ is analogous to a complex representation of the electric field envelope, the prime denotes differentiation with respect to the length variable $z$ and modes 3 and 4 are generalizations of what might conventionally be complex conjugates of modes 1 and 2, respectively. This somewhat abstract generalization to a four mode system enables the system to encompass both the Hermitian and PT two mode cases as special cases. This system conserves the following quantities, which are not to be confused with the definitions of similarly labelled quantities in section \ref{bquad}:
\begin{equation}\label{bEQ32}
\begin{aligned}
P=&A_{1}A_{3}+A_{2}A_{4}\\
K=&A_{1}A_{3}-A_{2}A_{4}-A_1A_2A_3A_4-\frac{1}{4}\left(A_1^2+A_2^2\right)\left(A_3^2+A_4^2\right)
\end{aligned}
\end{equation}
To solve this system we will largely follow the procedure in section \ref{bquad}. Let us first define the following function and constants:
\begin{equation}\label{bEQ33}
\begin{aligned}
\psi(z)&=p_1-2A_1(z)A_3(z)\\
p_1&=\frac{P^2}{12}-\frac{K}{3}-\frac{4}{3}+P\\
p_2&=\frac{P^2}{12}-\frac{K}{3}-\frac{4}{3}-P\\
\end{aligned}
\end{equation}
and it can then be shown from \eqref{bEQ31}-\eqref{bEQ33} that after some algebra:
\begin{equation}\label{bEQ34}
\begin{aligned}
\left(\psi'\right)^2&=4\psi^3-g_2\psi-g_3\\
g_{{2}}&=4\left(\,{p_{{1}}}^{2}+p_{{1}}p_{{2}}+{p_{{2}}}^{2}\right)
+48p_{{1}}+48p_{{2}}+128\\
g_3&=-4\left( p_{{1}}+p_{{2}}+4 \right)\left( p_{{1}}p_{{2}}-4p_{{1}}-4p_{{2}}-16 \right)
\end{aligned}
\end{equation}
and hence that $\psi$ is a Weierstrass elliptic function $\wp$ formed with elliptic invariants $g_2$ and $g_3$ \cite{WhitWat1915} such that:
\begin{equation}\label{bEQ35}
\begin{aligned}
&\psi(z)=\wp(z-z_0)\\
&\left\{z_0\in\mathbb{C}:\wp(z_0)=\psi(0),\,\wp'(z_0)=-\psi'(0)\right\}
\end{aligned}
\end{equation}
where $\wp'$ is the derivative of the Weierstrass $\wp$ function. Let us also define the points $\xi_{j}\in\mathbb{C}$ such that:
\begin{equation}\label{bEQ36}
\begin{aligned}
&\left\{\xi_1:\wp(\xi_1)=p_1,\,\wp'(\xi_1)=i4\left(2p_1+p_2+4\right)\right\} \\
&\left\{\xi_2:\wp(\xi_2)=p_2,\,\wp'(\xi_2)=-i4\left(p_1+2p_2+4\right)\right\} \\
\end{aligned}
\end{equation}
and it can also be shown that $\wp'\left(\xi_1-\xi_2\right)=0$ and hence $\xi_1-\xi_2$ is equal to one of the half-periods $w_j$ (modulo lattice periods) \cite{WhitWat1915}. The values of $z_0$ and $\xi_{j}$ can again be determined numerically by inverting $\wp$ in a mathematical software package such as Maple; $\wp'$ determines the sign, and these points are uniquely defined modulo lattice periods of the doubly periodic $\wp$ function (\textit{see} \eqref{A1}). Let us define $F_1=F_3=A_1A_3$ and $F_2=F_4=A_2A_4$ to be functions analogous to the power evolution in modes 1 and 2, respectively. It then follows from \eqref{bEQ32}-\eqref{bEQ36} that:
\begin{equation}\label{bEQ37}
\begin{aligned}
F_j(z)&=\left(-1\right)^{j-1}\frac{1}{2}\left[\wp(\xi_j)-\wp(z-z_0)\right]\\
\end{aligned}
\end{equation}
where $j=1-4$. While \eqref{bEQ37} is more compact, it is again possible to avoid having to invert $\wp$ to find $z_0$ by following the same procedure in section \ref{bquad}. But notably, having obtained \eqref{bEQ37} one can then construct the logarithmic derivatives of $A_{j}$ in terms of $\wp,\wp'$. To do this, it is convenient to initially express the logarithmic derivatives of $A_{j}$ in terms of $F_j$ and its derivative $F_j'$ using \eqref{bEQ31}, then one can use \eqref{bEQ36} and \eqref{bEQ37} to obtain:
\begin{equation}\label{bEQ38}
\begin{aligned}
\frac{A_j'}{A_j}&=\frac{i}{2}b_j+\frac{1}{2}\frac{\wp'(\xi_{j})-F_j'}{F_j}\\
&=\frac{i}{2}b_j+\frac{1}{2}\frac{\wp'(\xi_{j})-\wp'(z-z_0)}{\wp(\xi_{j})-\wp(z-z_0)}\\
&=\frac{i}{2}b_j+\zeta(z-z_0+\xi_j)-\zeta(z-z_0)-\zeta(\xi_j)
\end{aligned}
\end{equation}
where $b_1=-\left(P/2+1\right)$, $b_2=-\left(P/2-1\right)$, $b_3=-b_1$, $b_4=-b_2$, $\xi_3=-\xi_1$, $\xi_4=-\xi_2$, $\zeta$ is the Weierstrass zeta function \cite{WhitWat1915} and the $\zeta$ form follows from known function identities (\textit{see} \eqref{B2}). \eqref{bEQ38} is then easily integrated and expressed in terms of Weierstrass $\sigma$ functions using \eqref{EQ9} to yield the second main result of this paper; the general complex envelope solutions to cubic nonlinear systems in the form:
\begin{equation}\label{bEQ39}
\begin{aligned}
A_{j}(z)&=c_j\frac{\sigma(z-z_0+\xi_j)}{\sigma(z-z_0)}e^{\rho_j z}\\
\rho_j&=\frac{i}{2}b_j-\zeta(\xi_j)
\end{aligned}
\end{equation}
where the $c_j$ are the integration constants and can be fixed by the initial conditions. The $\sigma$ functions are also expressible in terms of the Jacobi $\theta_1$ functions (\textit{see} \eqref{B4}) and again it is also possible to obtain a Kronecker double series \cite{Weil1976,WhitWat1915} for \eqref{bEQ39} which we present in \ref{AppK}.

\begin{figure}[ht]
\centering
  \begin{subfigure}{0.49\columnwidth}
  \includegraphics[width=1\columnwidth]{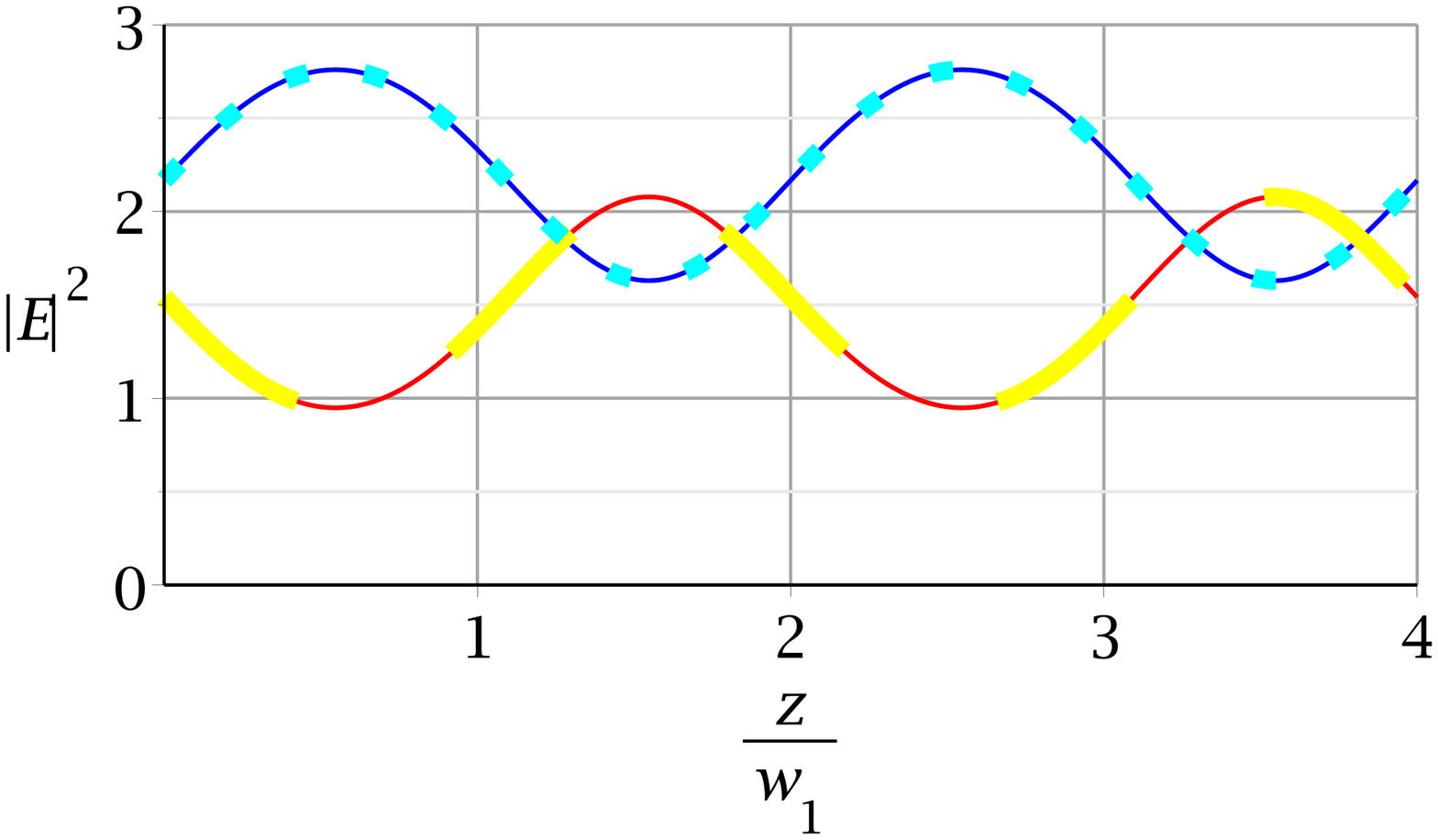}
	\caption{}
	\label{bfig1a}%
	\end{subfigure}
  \begin{subfigure}{0.49\columnwidth}
  \includegraphics[width=1\columnwidth]{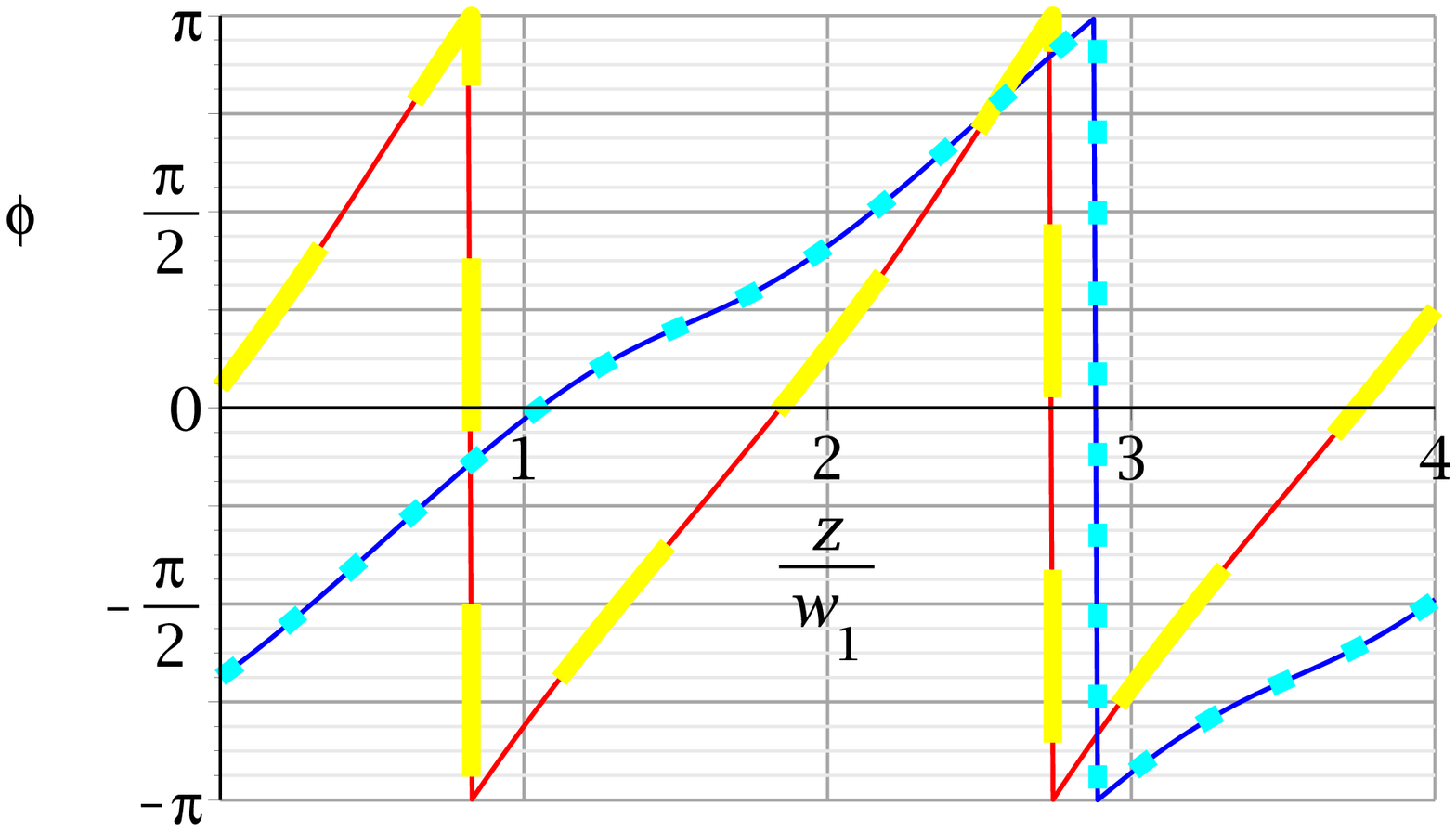}
	\caption{}
	\label{bfig1b}%
	\end{subfigure}
  \caption{(a) Power $|E|^2$ and (b) phase $\phi=\text{arg}(E)$ of $E_1$ (\eqref{bEQ39}: dash yellow, Numeric: solid red) and $E_2$ (\eqref{bEQ39}: dot cyan, Numeric: dark blue solid).}\label{bfig1}
\end{figure}	

Let us now consider two special examples of \eqref{bEQ31} from the literature. Firstly a Hermitian case can be obtained by setting:
\begin{equation}\label{bEQ40}
\begin{aligned}
A_1&=\sqrt{\frac{2}{3}}\,E_1\,e^{-2iPz}\\
A_2&=i\sqrt{\frac{2}{3}}\,E_2\,e^{-2iPz}\\
\end{aligned}
\end{equation}
where $P$ is as specified in \eqref{bEQ32}, and after imposing the additional constraints $A_3=A_1^*,\,A_4=A_2^*$, where the bar denotes complex conjugation, \eqref{bEQ31} becomes:
\begin{equation}\label{bEQ41}
\begin{aligned}
E_{1}'&=i E_{1}+i\left(|E_{1}|^2+\frac{2}{3}|E_{2}|^2\right)E_{1}+\frac{i}{3}\,E_2^2\,E_1^*\\
E_{2}'&=-i E_{2}+i\left(|E_{2}|^2+\frac{2}{3}|E_{1}|^2\right)E_{2}+\frac{i}{3}\,E_1^2\,E_2^*\\
\end{aligned}
\end{equation}
This system conserves the total power $|E_1|^2+|E_2|^2=\frac{3}{2}P$. \eqref{bEQ41} is, up to a scaling of the fields and the length variable, equivalent to the easily recognisable system describing the nonlinear interaction between two modes via the Kerr nonlinearity; for example $E_1$ and $E_2$ might describe continuous wave propagation in the two linearly polarised, phase velocity mismatched components of the electric field of an optical fibre \cite{Agrawal2007} (\textit{see} \ref{AppTrans} for how to transform from a conventional set of equations with material parameters in physical units to the normalised form in \eqref{bEQ41}). For the first time it is now evident that the general solution of this classic system is simply expressible via \eqref{bEQ39}. To obtain the solution we simply substitute \eqref{bEQ39} into \eqref{bEQ40} and express the parameters in \eqref{bEQ32}-\eqref{bEQ36} in terms of $E_1(0)$ and $E_2(0)$ using \eqref{bEQ40}. As a test, Fig. \ref{bfig1a} plots the magnitude and Fig. \ref{bfig1b} the phase of the analytic solution given in \eqref{bEQ39} for arbitrary launch conditions $E_1(0)=1.22+0.2\,i$ and $E_2(0)=-0.82-1.22\,i$ (dash yellow and dot cyan lines, respectively) together with numeric solutions of \eqref{bEQ31} obtained using a numerical Runge Kutta method (red and dark blue solid lines, respectively). The x-axis has been expressed in units of the half-period of oscillation $w_1$ which can be obtained via a standard elliptic integral involving the invariants in \eqref{bEQ34} (\textit{see} \eqref{A1}). The results are indistinguishable and together with more plots not presented, this supports the validity of the solutions. Physically the periodic exchange of power between modes in Fig. \ref{bfig1a} is indicative of the classic polarization rotation in nonlinear fibres and as there is no point along the length at which the power drops to zero in either mode, the polarization state is never linearly polarized in this example but rather it is always elliptically polarized.

\begin{figure}[ht]
\centering
  \begin{subfigure}{0.49\textwidth}
  \includegraphics[width=1\columnwidth]{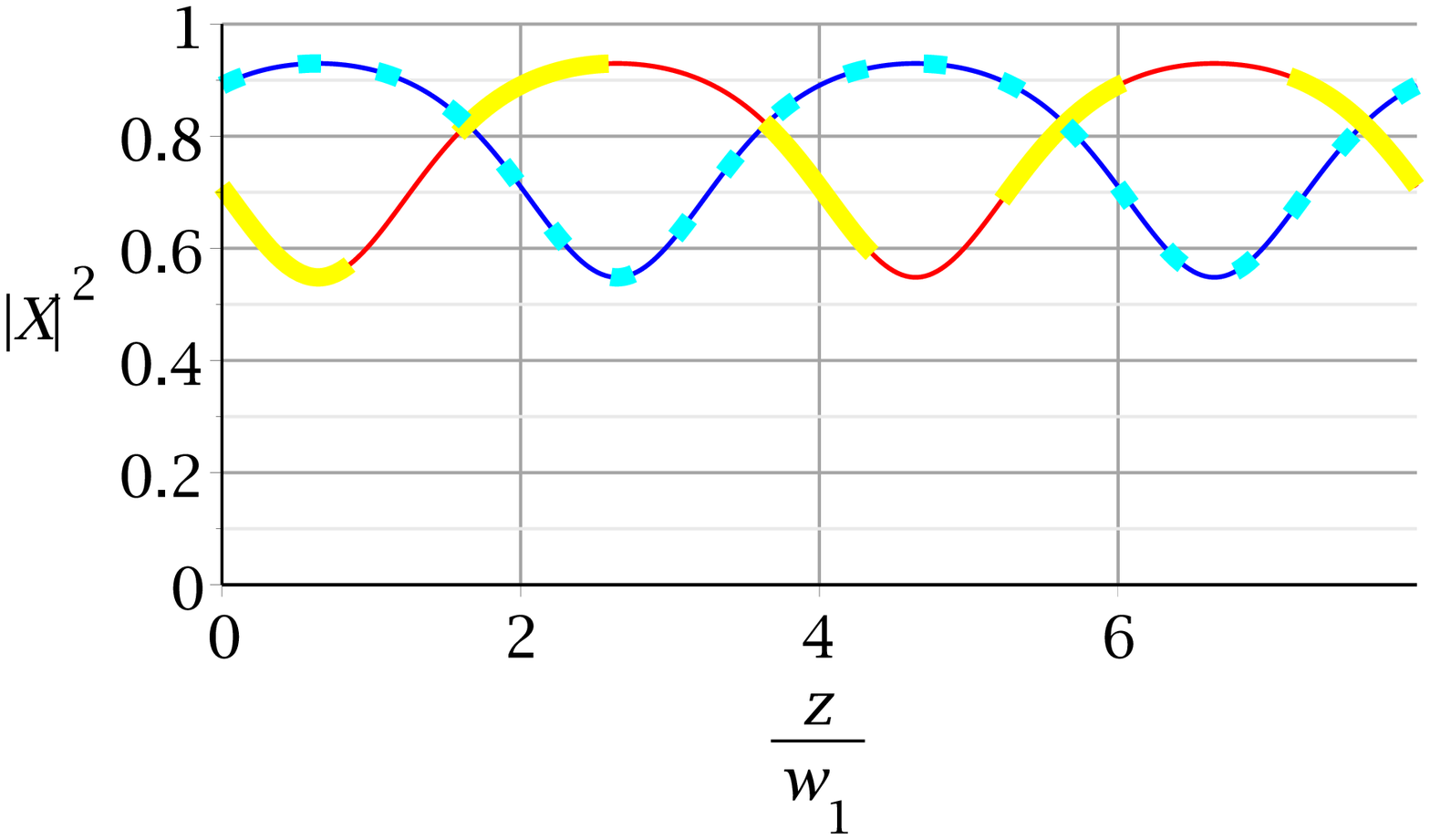}
	\caption{}
	\label{bfig2ab}%
	\end{subfigure}
  \begin{subfigure}{0.49\textwidth}
  \includegraphics[width=1\columnwidth]{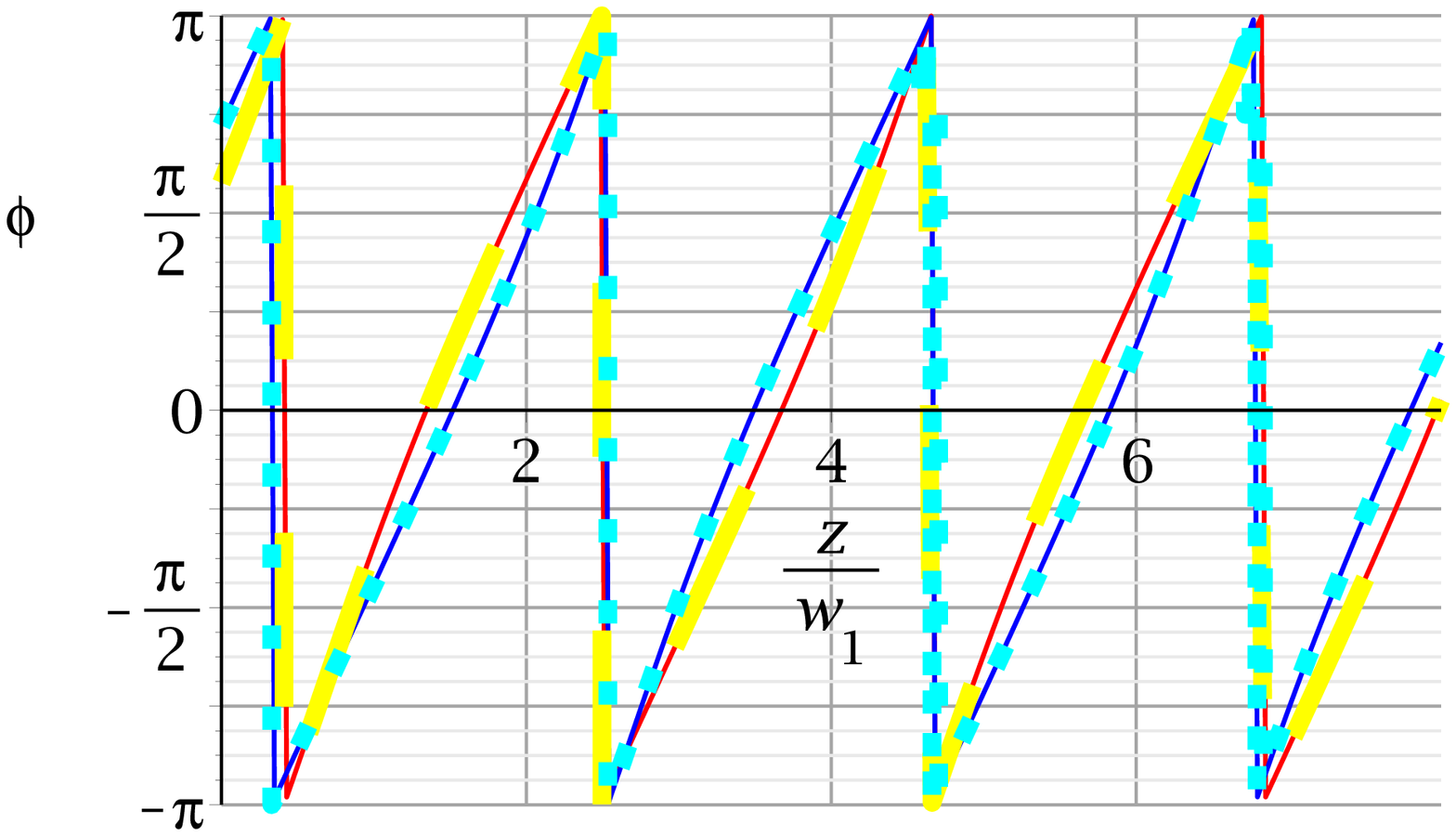}
	\caption{}
	\label{bfig2bb}%
	\end{subfigure}
  \caption{(a) Power $|X|^2$ and (b) phase $\phi=\text{arg}(X)$ of $X_1$ (\eqref{bEQ39}: dash yellow, Numeric: red solid)  and $X_2$ (\eqref{bEQ39}: dot cyan, Numeric: dark blue solid).}
\end{figure}

\begin{figure}[ht]
\centering
  \begin{subfigure}{0.49\textwidth}
  \includegraphics[width=1\columnwidth]{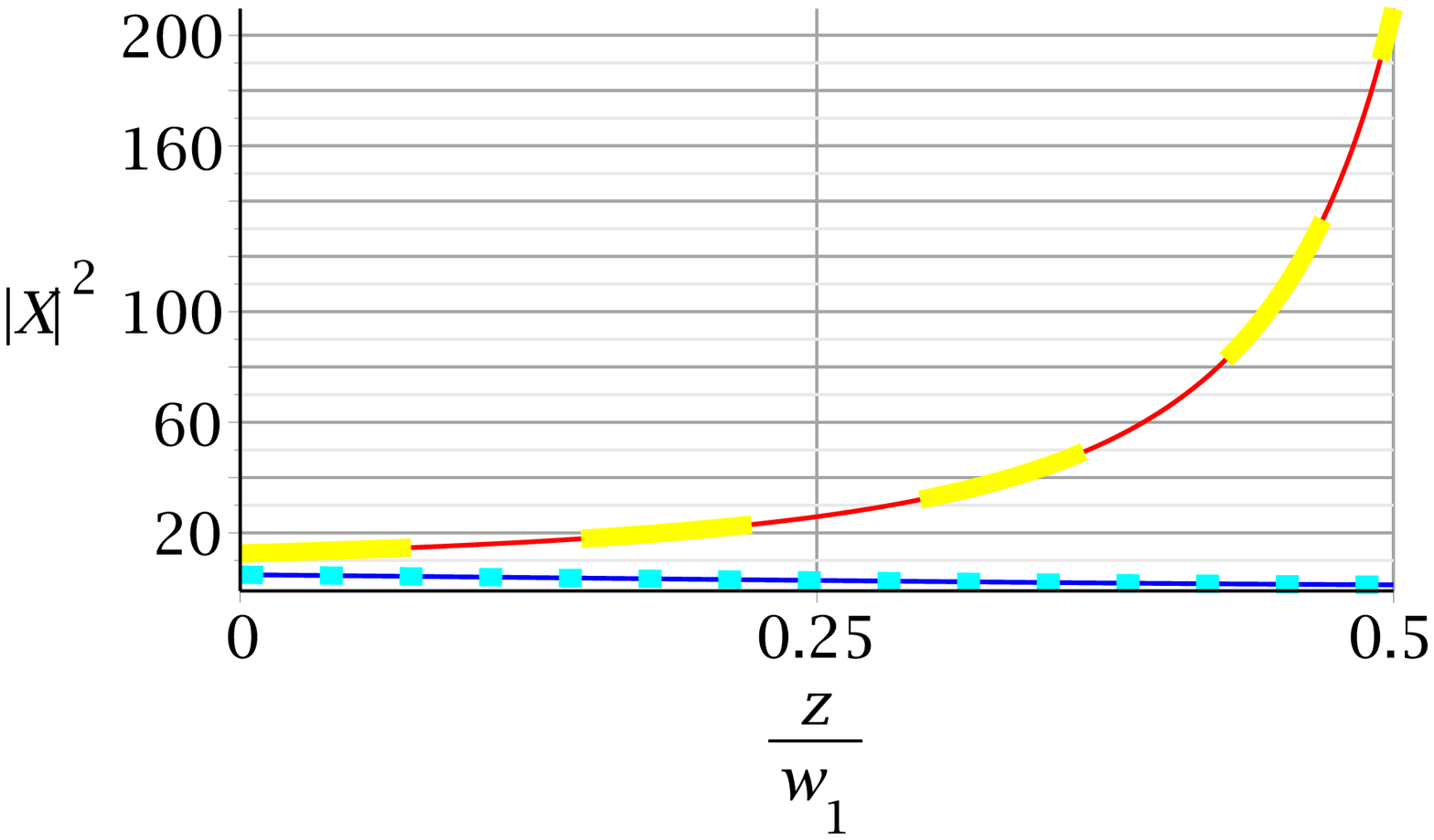}
	\caption{}
	\label{bfig2a}%
	\end{subfigure}
  \begin{subfigure}{0.49\textwidth}
  \includegraphics[width=1\columnwidth]{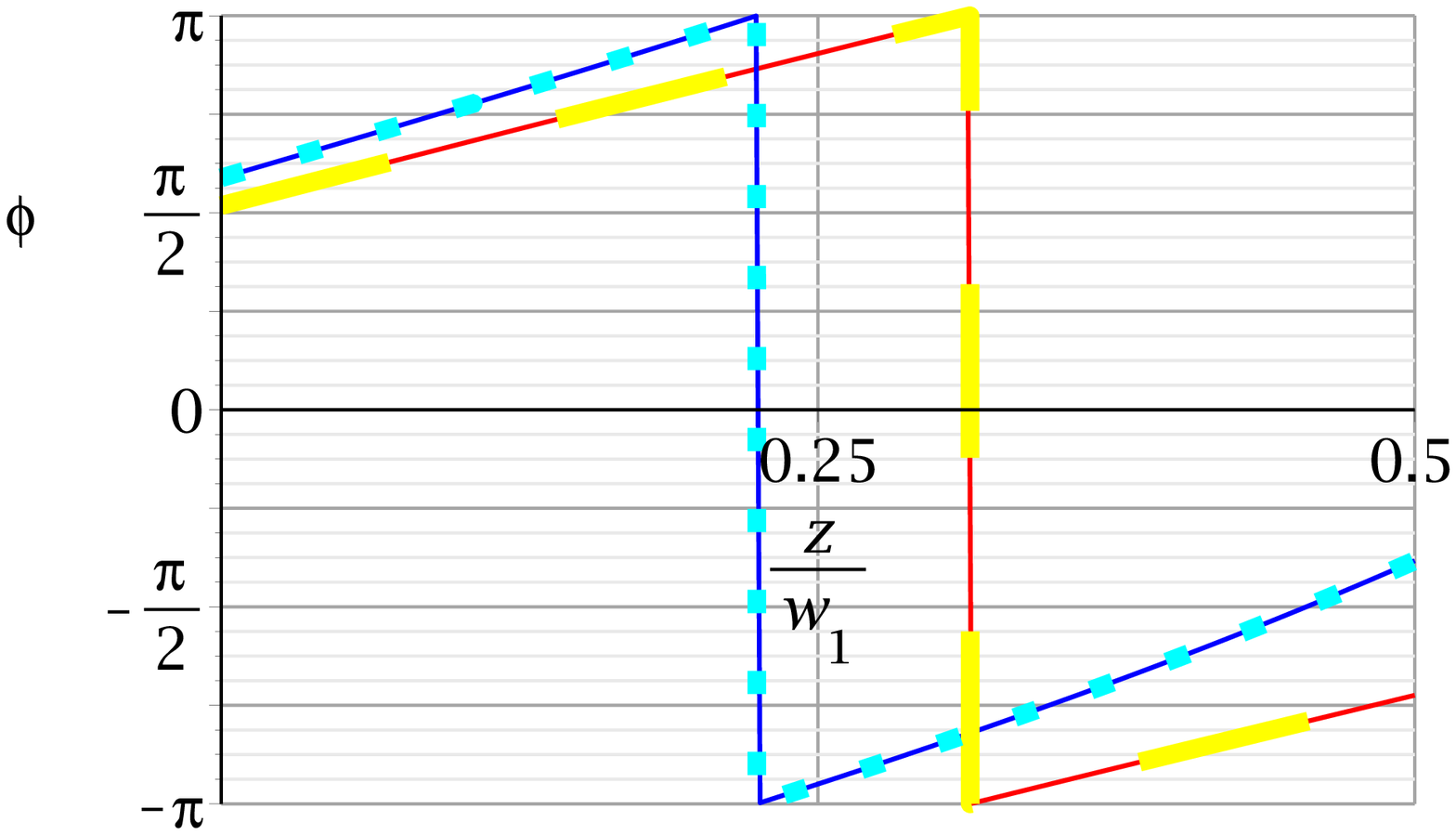}
	\caption{}
	\label{bfig2b}%
	\end{subfigure}
  \caption{(a) Power $|X|^2$ and (b) phase $\phi=\text{arg}(X)$ of $X_1$  (\eqref{EQ39}: dash yellow, Numeric: red solid)  and $X_2$ (\eqref{EQ39}: dot cyan, Numeric: dark blue solid).}\label{bfig2}
\end{figure}

The second case we will consider is the non-Hermitian PT case from \cite{Sarma2014} and can be obtained by setting:
\begin{equation}\label{bEQ11a}
\begin{aligned}
X_{1}&=\frac{i}{\sqrt{2}}\left(A_1+A_2\right)\\
X_{2}&=\frac{i}{\sqrt{2}}\left(A_1-A_2\right)\\
\end{aligned}
\end{equation}
together with the reductions $A_3=-A_1^*,\,A_4=A_2^*$, such that  \eqref{bEQ31} becomes:
\begin{equation}\label{bEQ11}
\begin{aligned}
X_{1}'&=i X_{2}+iX_{1}^2X_2^*\\
X_{2}'&=i X_{1}+iX_{2}^2X_1^*\\
\end{aligned}
\end{equation}
It is the minus sign in the relation between $A_3$ and $A_1^*$ that makes this system fundamentally different from a Hermitian case. This system, which was recently introduced as a discrete PT symmetric nonlinear nonlocal system, does not conserve the total power but instead it conserves the constant $P=X_{1}X_2^*+X_1^*X_{2}$. For more details of the physical realisation of this system and its properties the reader is referred to the original paper \cite{Sarma2014}, but it suffices to say here that this is a special case of \eqref{bEQ31} and thus the general solution is given by \eqref{bEQ39} together with the transformation in \eqref{bEQ11a}. One notable difference between this system and the Hermitian case is that the solutions can either be periodic or explosive depending on whether $\Delta=\left(1-|X_1\left(0\right)|^2\right)\left(1-|X_2\left(0\right)|^2\right)$ is positive or negative, respectively \cite{Sarma2014}. For example, Fig. \ref{bfig2ab} plots the magnitude and Fig. \ref{bfig2bb} the phase of the analytic solution given in \eqref{bEQ39} (dash yellow and dot cyan lines, respectively) for two modes with arbitrary launch conditions $X_1(0)=-0.2+0.81\,i$ and $X_2(0)=-0.6+0.71\,i$ $\left(\Delta>0\right)$ together with numeric solutions of \eqref{bEQ31} obtained using a numerical Runge Kutta method (red and dark blue solid lines, respectively). This example can be seen to be periodic; note though, that the period is $4w_1$ rather than $2w_1$ and this is due to the product term between $A_1$ and $A_2^*$ which appears when taking the magnitude of $X_1$ or $X_2$ in \eqref{bEQ11a}. In fact it can be shown that this term changes sign after a shift of $2w_1$. In contrast, Fig. \ref{bfig2a} plots the magnitude and Fig. \ref{bfig2b} the phase of the analytic solution given in \eqref{bEQ39} (dash yellow and dot cyan lines, respectively) for the launch conditions $X_1(0)=-0.2+2.5\,i$ and $X_2(0)=-0.61+2.12\,i$ $\left(\Delta<0\right)$ together with numeric solutions of \eqref{bEQ31} obtained using a numerical Runge Kutta method (red and dark blue solid lines, respectively). Here, the amplitude of $X_1$ can be seen to blow up on the approach to a pole while that of $X_2$ tends to zero. In all figures the analytic and numeric results are indistinguishable and together with more plots not presented, this again supports the validity of the solutions.  

\section{Wider Applications} \label{wideapps}

In the quadratic case, the form of the solution presented herein may prove particularly useful when trying to extend the algebraic manipulations to solitonic like or continuous wave solutions of the relevant time varying equations with dispersion \cite{Conforti2014, Zhang2015}. The methods presented herein may also generalize to higher mode systems such as waveguide arrays with quadratic nonlinearity \cite{Kobyakov1999, Setzpfandt2009, Setzpfandt2011}. In the cubic case, the solution method may prove particularly useful when trying to extend the algebraic manipulations to spatial or temporal solitonic like solutions of the relevant partial differential equations with diffraction or dispersion \cite{Sarma2011,Li2015} and it may also generalize to higher mode systems such as three-wave mixing \cite{Cappellini1991} and four-wave mixing \cite{Steffensen2011, Marhic2013}. As an extension outside of optics, the solution methods may also be applicable to Bose-Einstein condensate (BEC) models which combine quadratic and cubic nonlinearities \cite{Alexander2002}.

\section{Conclusions}

In summary, the general complex envelope solutions to the coupled mode systems which describe many optical processing functions and involve quadratic $\chi_2$ or cubic $\chi_3$ type nonlinearity, were derived for the first time. The solutions are expressible as a ratio of two Weierstrass $\sigma$ functions or as a Fourier series. Addition identities can mitigate the need to invert the elliptic functions in the amplitude solutions which will help display parameter dependence in physical problems. The analytic solutions were checked against numerical solutions obtained with a Runge Kutta method.

EPSRC grant EP/I01196X, The Photonics Hyperhighway. EPSRC Doctoral Prize.

The author would like to thank Periklis Petropoulos for a careful read of the manuscript.

\appendix
\begin{appendices}
\section{Rescaling Transformations} \label{AppTrans}
\eqref{At1} is a textbook model of the complex electric field $v_j$ at frequency $f_j$; with $f_3=f_1+f_2$, during continuous wave (stationary regime) sum- and difference-frequency generation in a quadratic nonlinear medium \cite{Trager2007} :
\begin{equation}\label{At1}
\begin{aligned}
v'_1(Z)&=i\gamma_1\, v_2^*\,v_3\, e^{-i\Delta k Z}\\
v'_2(Z)&=i\gamma_2\, v_1^*\,v_3 \, e^{-i\Delta k Z}\\
v'_3(Z)&=i\gamma_3\, v_1\,v_2\, e^{i\Delta k Z}\\
\end{aligned}
\end{equation}
where $\Delta k=k_1+k_2-k_3$ is the wave vector mismatch and $\gamma_j$ is a positive real frequency dependant nonlinear coefficient. The transformation:
\begin{equation}\label{At2}
\begin{aligned}
v_j(Z)&=\Delta k\sqrt{\frac{s_j\, \gamma_j}{\gamma_1\gamma_2\gamma_3}}\,a_j(Z) \, e^{-i\Delta k Z}\\
\end{aligned}
\end{equation}
where $s_1=s_2=2$ and $s_3=1$, followed by the introduction of a new length variable $z=\Delta k Z$ and function $A_j(z)=a_j(Z)$ yields \eqref{EQ31}.

\eqref{At3} is a textbook model of the two orthogonal linear polarization modes of an optical fibre $u_j$ in the continuous wave (stationary regime) \cite{Agrawal2007}:
\begin{equation}\label{At3}
\begin{aligned}
u_{x}'(Z)&=i\gamma\left(|u_{x}|^2+\frac{2}{3}|u_{y}|^2\right)u_{x}+\frac{i\gamma}{3}\,u_y^2\,u_x^*\,e^{-2i\Delta \beta Z}\\
u_{y}'(Z)&=i\gamma\left(|u_{y}|^2+\frac{2}{3}|u_{x}|^2\right)u_{y}+\frac{i\gamma}{3}\,u_x^2\,u_y^*\,e^{2i\Delta \beta Z}\\
\end{aligned}
\end{equation}
where $\Delta \beta=\beta_{0x}-\beta_{0y}$ is the phase velocity mismatch and $\gamma$ is the nonlinear coefficient. The transformation:
\begin{equation}\label{At4}
\begin{aligned}
u_x(Z)&=\sqrt{\frac{\Delta \beta}{2\gamma}}\,a_1(Z) \, e^{-i\frac{1}{2}\Delta \beta Z}\\
u_y(Z)&=\sqrt{\frac{\Delta \beta}{2\gamma}}\,a_2(Z) \, e^{i\frac{1}{2}\Delta \beta Z}\\
\end{aligned}
\end{equation}
followed by the introduction of a new length variable $z=2\Delta \beta Z$ and function $E_j(z)=a_j(Z)$ yields \eqref{bEQ41}.

\section{Periods as elliptic integrals} \label{AppA}
The power evolution of the modes is periodic and, in the case of real roots $e_1>e_2>e_3$ of $4x^3-g_2x-g_3$, the real period of oscillation is given by $2w_1$ where:
\begin{equation}\label{A1}
\begin{aligned}
w_1=\int_{e_1}^{\infty}\frac{1}{\sqrt{4x^3-g_2x-g_3}}{dx}
\end{aligned}
\end{equation}
while the second imaginary half period is defined by:
\begin{equation}\label{A2}
\begin{aligned}
w_3=-i\int_{-\infty}^{e_3}\frac{1}{\sqrt{4x^3-g_2x-g_3}}{dx}
\end{aligned}
\end{equation}
Together these generate the lattice of the doubly-periodic $\wp$ function. In the cubic case in section \ref{bcube}, the roots themselves can be explicitly given by:
\begin{equation}\label{A3}
\begin{aligned}
-\frac{4}{3}-\frac{{P}^{2}}{6}-\frac{2\,K}{3},\quad
\frac{{P}^{2}}{12}+\frac{K}{3}+\frac{2}{3}\pm\sqrt{2\,{P}^{2}+4\,K+4}
\end{aligned}
\end{equation}

\section{Elliptic function identities} \label{AppB}
\begin{equation}\label{B1}
\wp(x+y)=\frac{1}{4}\left(\frac{\wp(x)'-\wp(y)'}{\wp(x)-\wp(y)}\right)^2-\wp(x)-\wp(y)
\end{equation}
\begin{equation}\label{B2}
\frac{1}{2}\frac{\wp(x)'-\wp(y)'}{\wp(x)-\wp(y)}=\zeta(x+y)-\zeta(x)-\zeta(y)
\end{equation}
\begin{equation}\label{B3}
\wp(x)-\wp(y)=-\frac{\sigma(x+y)\sigma(x-y)}{\sigma(x)^2\sigma(y)^2}\\
\end{equation}
\begin{equation}\label{B4}
\begin{aligned}
\\
\sigma(z)&=\frac{2w_1}{\pi}\exp{\left[\frac{\zeta(w_1)z^2}{2w_1}\right]}\frac{\theta_1(\frac{\pi z}{2w_1},q)}{\theta_1'(0,q)}\\
q&=e^{i\pi\tau},\,\tau=w_3/w_1
\end{aligned}
\end{equation}

\section{Kronecker Double Series} \label{AppK}
Variants of the following double series expression are attributed to Kronecker \cite{Weil1976,WhitWat1915}:
\begin{equation}\label{K1}
\begin{aligned}
\kappa(x,y)=&\frac{\sigma(x+y)}{\sigma(x)\sigma(y)}e^{-x y\,\frac{\zeta(w_1)}{w_1}}\\
=&\frac{\pi}{2w_1}\left[\cot\left(\frac{\pi x}{2 w_1}\right)+\cot\left(\frac{\pi y}{2 w_1}\right)\right]\\
&+\frac{2\pi}{w_1}\sum_{m=1}^{\infty}\sum_{n=1}^{\infty}\sin\left(\frac{\pi}{ w_1}\left[mx+ny\right]\right)e^{2i\pi\, m\, n\, \tau}
\end{aligned}
\end{equation}
which converges absolutely provided that $\left|\mathbb{I} [x/\left(2\,w_1\right)]\right|<\left|\mathbb{I} [\tau]\right|$ and $\left|\mathbb{I} [y/\left(2\,w_1\right)]\right|<\left|\mathbb{I} [\tau]\right|$ where $\left|\mathbb{I} [\cdot]\right|$ denotes the magnitude of the imaginary part.
\eqref{EQ39} and \eqref{bEQ39} can be expressed as a Kronecker double series by noting that:
\begin{equation}\label{K2}
\begin{aligned}
A_{j}(z)&=c_j\,\kappa(z-z_0,\xi_j)\,\sigma(\xi_j)\,e^{\left(z-z_0\right)\,\xi_j\,\frac{\zeta(w_1)}{w_1}}e^{\,\rho_j \,z}\\
\end{aligned}
\end{equation}

\end{appendices}

\end{document}